\documentclass[a4paper,11pt]{article}
\pdfoutput=1 
\usepackage{jcappub} 
\usepackage[T1]{fontenc}
 \usepackage{amssymb,amsfonts,slashed,amsthm,amsmath,graphicx}
\usepackage[all]{xy}
\usepackage{dcolumn}
\usepackage{bm}
\usepackage{mathrsfs}
\usepackage{ucs}
\usepackage[utf8x]{inputenc}
\usepackage{hyperref}
\usepackage{hepunits}
\usepackage{slashed}
\usepackage{floatrow}
\usepackage{subfigure}
\usepackage{indentfirst}
\usepackage{comment}
\usepackage{multicol}
\begin{document}

\newcommand{\be}{\begin{equation}}
\newcommand{\ee}{\end{equation}}
\newcommand{\bq}{\begin{eqnarray}}
\newcommand{\eq}{\end{eqnarray}}
\newcommand{\bsq}{\begin{subequations}}
\newcommand{\esq}{\end{subequations}}
\newcommand{\bc}{\begin{center}}
\newcommand{\ec}{\end{center}}
\bibliographystyle{JHEP}

\title{\boldmath {\LARGE{Measuring Global Monopole Velocities, one by one\vspace{-2.5 cm}}}}

\affiliation[a]{Department of Theoretical Physics, University of the Basque Country UPV/EHU,\\
48080 Bilbao, Spain}
\author[a]{{\large Asier Lopez-Eiguren,}}
\author[a]{{\large Jon Urrestilla}}
\author[a,b]{{\large and Ana Ach\'ucarro}}
\affiliation[b]{Institute Lorentz of Theoretical Physics, University of Leiden, 2333CA Leiden, The Netherlands}

\emailAdd{asier.lopez@ehu.eus}
\emailAdd{jon.urrestilla@ehu.eus}
\emailAdd{achucar@lorentz.leidenuniv.nl}

\abstract{
We present an estimation of the average velocity of a network of global monopoles in a cosmological setting using large numerical simulations. In order to obtain the value of the velocity, we improve some already known methods, and present a new one. This new method  estimates individual global monopole velocities in a network, by means of detecting each monopole position in the lattice and following the path described by each one of them. Using our new estimate we can settle an open question previously posed in the literature:  velocity-dependent one-scale (VOS) models for global monopoles predict two branches of scaling solutions, one with monopoles moving at subluminal speeds and one with monopoles moving at luminal speeds. Previous attempts to estimate monopole velocities had large uncertainties and were not able to settle that question.  Our simulations find  no evidence of a luminal branch.  We also estimate the values of the parameters of the VOS model. With our new method we can also study the microphysics of the complicated dynamics of individual monopoles.  Finally we use  our large simulation volume to compare the results from the different estimator methods, as well as to asses the validity of the numerical approximations made.
}

\maketitle
\flushbottom

\section{Introduction}
\label{intro}
In his pioneering work, Kibble \cite{Kibble:1976sj} showed that if topological defects are stable, they are necessarily formed at cosmological phase transitions. The properties of the topological defects created will depend on details of the phase transition that created them, and in particular, on the specific symmetry being broken and the energy scale at which the transition happened. Understanding the  formation, dynamical properties and evolution of defects is therefore a key part of any effort to understand the physics governing the early universe. 

	However, defect evolution is a difficult problem involving physics from very different scales: defects are formed in the very early universe and evolve and survive until much later in the history of the universe; even to today.  In order to deal with these difficulties two main complementary techniques have been used traditionally to obtain the properties of defect networks: numerical simulations and  analytic, phenomenological models. On the one hand, numerical simulations resolve the equations of motion of a given defect type, under different levels of approximation. They require a high computational cost and have a limited accuracy and a limited evolution time. On the other hand, analytic models are effective models that capture the properties of a network of defects into a simpler and more tractable set of evolution equations for a small number of physically meaningful macroscopic quantities that describe the network, such as typical velocities or length scales, 
in terms of a few phenomenological parameters that need be determined (calibrated) by  numerical simulations of the  true microphysics of the system. The symbiosis of those approaches is clear then: effective models need input from numerical simulations; but once the effective models have been successfully calibrated, they are much more tractable and make predictions easier. One could sometimes also revert to numerical simulations in regions of parameter space that the effective model has pinpointed as interesting or relevant.

	The first analytical study of topological defects networks was made by Kibble \cite{Kibble:1984hp} for cosmic strings. In Kibble's model, known as  the 'one-scale' model, the evolution of long-string segments is described using a single meaningful macroscopic quantity, a length-scale which is usually called the 'correlation length'. Solutions for this model were the first ones to show the existence and stability of scaling solutions, subject to conditions on the loop production mechanism, which would be later confirmed by numerical simulations of string network evolution \cite{Bennett:1989yp,Allen:1990tv}. In particular, these numerical simulations showed  that small scale structure takes a crucial role in the network dynamics. Simulations revealed the existence of a significant amount of small-scale structure on long strings, with loops being predominantly produced at the smallest scales that can be sampled numerically. 

	Based on these findings, Austin, Copeland and Kibble developed a model \cite{Austin:1993rg} in order to account for the small-scale structure. This model made use of three different length scales: the length-scale used by Kibble, which can be understood as the inter-defect distance; a length-scale which explicitly describes the small structure; and a length-scale which introduces the effects of gravitational radiation. This model confirmed the predictions of the one-scale  model for the large-scale properties of the network.  

	After these attempts to build a model to describe the evolution of string networks, Martins and Shellard realised that it would be relevant to take into account frictional forces due to particle-string scattering, which are important for some time after the string-forming phase transition. The model they proposed, known as  the 'velocity-dependent one-scale' (VOS) model \cite{Martins:1996jp, Martins:2000cs}, is a simple generalisation of the  one-scale model where the average  velocity (root-mean-square velocity) of the string network becomes a dynamical variable. The description of the evolution of cosmic string networks that can be obtained using this model is a fully quantitative description of its complete evolution in the early universe, although  the small-structure and other potentially important effects such as loop reconnections are not included. 
	
	Encouraged by the success, and the simplicity, of this model to describe the evolution of string networks, the same procedure was used to describe different topological defect network evolutions. In fact, similar procedures have been used to describe the evolution of domain-walls \cite{Avelino:2005kn,Martins:2016ois}, of  monopoles \cite{Martins:2008zz} and of  semilocal-strings \cite{Nunes:2011sf,Achucarro:2013mga,Lopez-Eiguren:2014xaa}; as well as of extended objects in superstring theory  \cite{Copeland:2005cy,Avgoustidis:2007aa,Sousa:2011iu,Avgoustidis:2014rqa}, just to cite a few.

The first contribution to the analytical description of monopole dynamics  was made  in \cite{Martins:2008zz}, where the authors  proposed an extension of the VOS model that can be used to study the evolution of local or global monopole networks. In this work they  found that for the case of local monopoles  the scaling solutions are quite robust. However,  for the case of global monopoles,  if the expansion rate is not too fast, the model admits  two different scaling solutions corresponding to two different velocity branches: $v<1$ (subluminal) and $v=1$ (luminal).

As mentioned before, input from numerical simulations is needed to determine the values of the parameters in the effective models. Most of the numerical simulations for defects have been centered around cosmic strings (see these recent works \cite{Ringeval:2012tk,Lazanu:2014xxa,Daverio:2015nva,Blanco-Pillado:2015ana,Charnock:2016nzm} and references therein as examples of different approaches) and string-like objects, closely related to cosmic strings \cite{Achucarro:2005tu,Hindmarsh:2006qn,Urrestilla:2007yw,Rajantie:2007hp,Lizarraga:2016hpd,Hindmarsh:2016dha}. 
	
With regard to monopoles, there have been some attempts to study their network evolution. In some papers (for example \cite{Bennett:1990xy,Perivolaropoulos:1991du,Durrer:1998rw}), the focus was on determining their effect on the Cosmic Microwave Background anisotropies. In another work by Yamaguchi \cite{Yamaguchi:2001xn}, the properties of the network were studied, and most importantly from  the point of view of the present work, the first attempt to estimate the velocities of the monopoles was made. The results for the average monopole network velocities in radiation--  and matter--dominated universes in \cite{Yamaguchi:2001xn} are $v_r=1.0\pm0.3$ and $v_m=0.8\pm0.3$ respectively. The methods used in that work had a rather large error; big enough that, for example,  one cannot determine whether the luminal or subluminal branches of the global monopole VOS model should be used.

 In this paper we measure the global monopole velocities using the largest and most accurate field theory simulations of these objects to date in order to determine more precisely if the global monopoles are luminal or subluminal, and to calibrate the analytical model proposed in \cite{Martins:2008zz}. In order to do so we develop a new method that determines the velocity of each one of the monopoles in the simulation box at every time-step.  We also implement   the method proposed by Yamaguchi \cite{Yamaguchi:2001xn}, and an extension of the method proposed by Hindmarsh et al. \cite{Hindmarsh:2008dw}. 
In doing so we will determine more accurately the values of the parameters for the monopole VOS model, and also give a measure of the accuracy of each velocity estimator, as well as   the accuracy of the numerical approximations used for the simulations.

These simulations are not only interesting for networks of global monopoles. In fact, global monopoles are also present in the semilocal string model \cite{Vachaspati:1991dz,Hindmarsh:1991jq,Achucarro:1999it}. These strings are non-topological, and thus can have ends. The field configuration around the ends of the strings could be identified with some sort of global monopole. Therefore, the new method proposed here can be used to calibrate the semilocal string analytic models proposed in \cite{Nunes:2011sf}.

The rest of the article is organised as follows. In Sect.~\ref{sec-ms} we introduce the global monopole model and  describe the field theory simulations that we performed. We then present  three different velocity estimators in  Sect.~\ref{tv}, including our new estimator. We report the results obtained from those methods in Sect.~\ref{re-v}. In Sect.~\ref{an-ca} we use the results obtained to calibrate the analytic model for global monopoles. Finally we conclude in Sect.~\ref{co}.

\section{Field Theory Simulations of Global Monopoles}
\label{sec-ms}

\subsection{Global Monopoles \label{sec-model}}

The simplest model that gives rise to global monopoles is described by the following action \cite{Barriola:1989hx},
\be
{\cal S}=\int d^4x \sqrt{-g} \Big[ \frac{1}{2} \partial_{\mu}\Phi^i \partial^{\mu}\Phi^i-\frac{1}{4}\lambda(|\Phi|^2-\eta^2)^2 \Big],
\label{action}
\ee
where $\Phi^i$, $i=1,2,3$ is a real scalar triplet, $|\Phi|\equiv \sqrt{\Phi^i\Phi^i}$ and $\lambda$ and $\eta$ are real constant parameters. Since our aim is to study the dynamics of a  network of global monopoles in a expanding universe, so we consider a   flat Friedmann-Robertson-Walker space-time with comoving coordinates:
\be
ds^2=a(t)^2\left[dt^2-dx^2-dy^2-dz^2\right]\,,
\ee
 where $a(t)$ is the cosmic scale factor and $t$ is conformal time.

The model has a global $O(3)$ symmetry spontaneously broken down to $O(2)$, leading to two Goldstone bosons and one scalar excitation with $m_s=\sqrt{2\lambda}\eta$. The set of ground states is the two sphere $|\Phi|=\eta$ and, since $\pi_2(S^2)={\bf Z}$, there are field configurations with non-trivial topological charge. For example, a global monopole of unit topological charge can be described by the "hedgehog" configuration $\phi^i=\phi(r)\hat{x}^i$, where $\hat{x}^a$ is a radial unit vector and outside the monopole core $\phi(r)\approx\eta$.  

The equations of motion derived from the action (\ref{action}) are 
\be
\ddot{\phi}^i+2\frac{\dot{a}}{a}\dot{\phi}^i-\nabla^2 \phi^i=-a^2\lambda(\phi^2-\eta^2)\phi^i,
\label{eom}
\ee 
and the dots represent  derivatives with respect to the conformal time $t$.

 At this point we can observe that the monopole size $\delta\approx (\sqrt{\lambda}\eta)^{-1}$, which is a fixed physical length scale, rapidly decreases in  comoving coordinates. Thus, in order to obtain a representative network, extreme care is needed when  setting the parameters controlling the evolution of the simulations. Otherwise the monopole size could be too small to be resolved during the simulation, or too large initially and monopole cores could be overlapping. This is a well-known issue in lattice simulations, and that difficult is overcome by the use of the Press-Ryden-Spergel algorithm \cite{Press:1989yh}. The algorithm proposes to turn the coupling constant into a time-dependent variable: 
\be
\lambda=\lambda_0 a^{-2(1-s)}\,,
\label{lambda-s}
\ee
where the parameter $s$ controls the sensitivity of the monopole size to the scale factor $a$. The value $s=0$ corresponds to  a fixed monopole size in comoving coordinates and the value $s=1$ corresponds to the true case.  Previous works using the  algorithm in cosmic string \cite{Moore:2001px,Bevis:2010gj,Daverio:2015nva,Lizarraga:2016onn} prove the validity of the algorithm.  The errors due to the algorithm  are typically smaller than the statistical errors, or the systematic errors inherent to the discretization procedure.   Performing the following rescalings
\begin{eqnarray}
\Phi^i &\rightarrow & \tilde{\Phi}^i=\frac{\Phi^i}{\eta}\,,
 \nonumber\\
x^{\mu} & \rightarrow & \tilde{x}^{\mu}=\sqrt{\lambda_0 \eta^2}x^{\mu}\,,
\end{eqnarray}
and substituting (\ref{lambda-s}) in (\ref{eom}),   we now have:
\be
\ddot{\phi}^i+2\frac{\dot{a}}{a}\dot{\phi}^i-\nabla^2 \phi^i=-a^{2s}(\phi^2-1)\phi^i.
\ee

Global monopoles are rather interesting objects intrinsically. 
The energy of a monopole is (linearly) divergent with radius, due to the slow fall-off of angular gradients. In a cosmological situation this divergence is not catastrophic, since there is always an antimonopole around that cuts-off the energy divergence.  The force between a well-separated monopole and anti-monopole is approximately independent of their distance \cite{Perivolaropoulos:1991du} and simulations often show how a monopole-antimonopole pair that are nearby 'repel' each other, since they have found another partner to annihilate with somewhere else.  The stability of global monopoles was also the subject of debate: if the core of the monopoles was artificially fixed, it was argued that they were unstable towards concentrating all the gradient energy in the, say, north pole, and then decaying into the vacuum \cite{Goldhaber:1989na}, pointing towards an instability\footnote{Global monopoles were   proven to be stable to axisymmetric normalizable perturbations \cite{Achucarro:2000td}, though (surprisingly) the energy barrier between different topological defects is {\it finite}.}. However, if the core of the monopole is free to move, it was argued in \cite{Rhie:1990kc} that the core would move upwards so as to compensate the increase of the concentration of gradients in the north pole, and therefore there is no instability, just the translation of the monopole core.

\subsection{Simulations and Scaling} 
Instead of solving directly the equations of motion, we discretise the action (\ref{action}), and obtain the equations of motion corresponding to the discretized actions. These are translated into a cartesian grid with standard techniques (see for example  \cite{Achucarro:2005tu}) and evolved in $2048^3$ lattices with periodic boundary conditions.  The simulations were parallelized using  the LatField2 library for parallel field theory simulations \cite{David:2015eya}. The simulations were performed at the COSMOS Consortium supercomputer and i2Basque academic network computing infrastructure.

Notice that there is an upper limit on the time that the system can be evolved before it feels the effects of the periodic boundary condition. The simulation can only be believed up to half light-crossing time, i.e., if we sent a light ray in opposite directions in the box, the simulation is accurate up to when the two rays meet again. Therefore there is a clear compromise when choosing the values for the space ($\Delta x$) and time ($\Delta t$) discretization in the lattice: finer lattice discretization would mean that the solution is more accurate, to the expense of having a smaller dynamical range; whereas if the discretization is coarser, the dynamical range increases but the equations are not be solved accurately enough.
As a good compromise, for the computing power available to us, we chose $\Delta x=0.5$ and $\Delta t=0.25$, where both $\Delta x$ and $\Delta t$ are measured in units of $[\eta^{-1}]$. 

We have simulated the system in  radiation and matter dominated eras, for two different values of the $s$ parameter: $s=0$ and $s=1$. We have performed five production simulations for each case  which, given the high number of monopoles in each simulations, give us appropriate statistics. 

The key property of a  global monopole network that allows us to study it by using numerical simulations is scaling, i.e.,  the regime where  the typical scales of the global monopole network depend linearly with the horizon size \cite{Bennett:1990xy,Yamaguchi:2001xn}. Therefore, it is desirable that our simulations reach scaling as fast as possible, and our choice of initial conditions is such that it enables the system to approach the scaling solution as fast as possible. Once scaling is reached the system forgets its initial configuration, therefore, the only importance of the initial configuration is to drive the system to scaling as fast as possible. This way we would have a long dynamical range and we could estimate more accurately monopole velocities.   We found that    Vachaspati-Vilenkin \cite{Vachaspati:1984dz} type initial conditions are  a good choice for our case: the scalar field velocities are set to zero and scalar fields are chosen to lie in the vacuum manifold, but have randomly chosen orientations. This field configuration has to relax into a network of scaling monopoles and we achieve this by using a period of diffusive evolution, with the second derivatives removed from the equations of motion. 

Depending on the nature of the simulation (cosmological era and value of $s$) the system needs to undergo different  periods of diffusion $t_{\rm dif}$  in order to aid in reaching the scaling regime. After the diffusion regime, the equations of motion are solved. The scaling regime is obtained at different stages in each case, and bearing that in mind, we start extracting data from time $t_{\rm ini}$ until $t_{\rm end}$. It is numerically very expensive to analyze the system every single time-step, but it needs to be analysed often enough to obtain meaningful results. The compromise between these two situations is reached by analyzing the system every $t_s$ time-units. The parameters and details of each case  can be found in table~\ref{params}.

\begin{table*}
\begin{center}
\begin{tabular}{ |c|c|c|c|c|c|}\hline
   $s$ & Cosmology &  $t_{\rm dif}$ & $t_{\rm ini}$ & $t_{\rm end}$ & $t_s$ \\\hline
0& Mat\&Rad &   12.5& 150& 510 & 5 for $t<200$, otherwise 10   \\  \hline
1 & Rad& 25 & 210 & 510 &  10 \\ \hline
1 & Mat & 25 & 250 & 510 & 10 \\\hline
\end{tabular}
\caption{\label{params} Description of the time parameters for every different simulation, as explained in the text.}
\end{center}
\end{table*}

\begin{figure}
\includegraphics[width=0.9\textwidth]{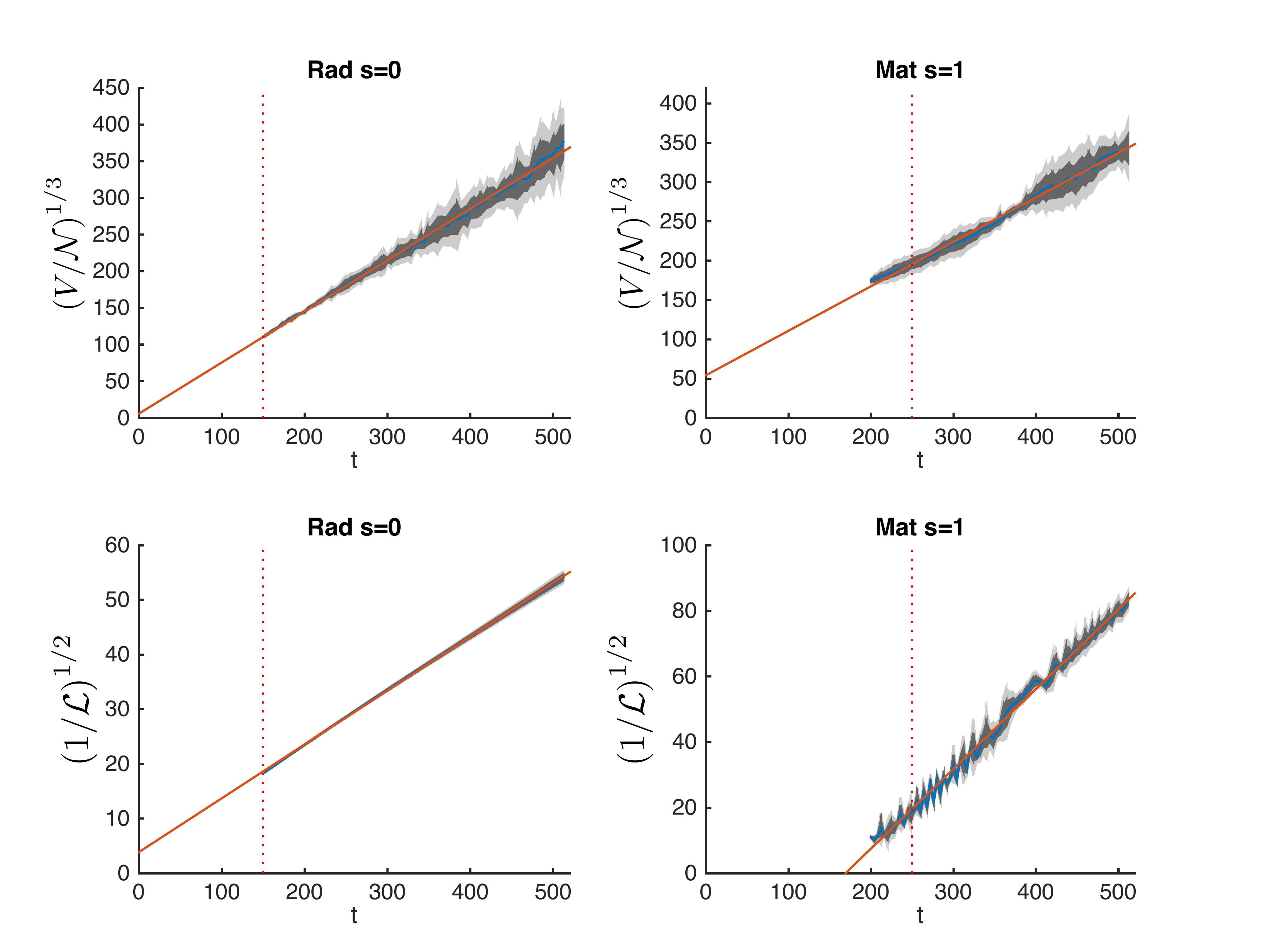}
\caption{Scaling regions computed using the monopole number ($\cal N$) and the Lagrangian density ($\cal L$) as the characteristic lengths of the network. The two extreme cases are shown: radiation era with $s=0$, where the system reaches a smooth scaling regime in a comparatively long period of time; and Matter era with $s=1$, where the scaling regime is not so smooth and in a shorter time interval.   }
\label{fig-sca}
\end{figure}

It is of great importance to test the scaling regime of the system before starting to acquire meaningful results. We test it by monitoring that a characteristic length of the network grows linearly in time. In fact,  we use two different  characteristic lengths in this work:

On the one hand, we can define a velocity-one-scale (VOS) type length-scale \cite{Martins:1996jp,Martins:2000cs}
\be
\Big(\frac{V}{{\cal N}}\Big)^{1/3}=\gamma_{m} t\,,
\label{gammat}
\ee 
where V is the volume of the simulation box, ${\cal N}$ is the number of monopoles\footnote{We will refer to monopoles  {\it and} antimonopoles as just monopoles, for simplicity, unless the distinction is meaningful, where we will revert to distinguishing them.}  in the simulation box, and $\gamma_m$ is a proportionality constant.  In figure~\ref{fig-sca}  we show that, after a burn in period,  the simulation reaches a regime where  $(V/{\cal N})^{1/3}$ is  approximately linear with respect to $t$.

On the other hand, we can obtain the characteristic length scale of the network from the Lagrangian \cite{Bevis:2006mj},
\be
\Big(\frac{\eta^2}{{\cal -L}}\Big)^{1/2}=\gamma_{{\cal L}} t\,,
\ee 
where $\gamma_{{\cal L}}$ is another proportionality constant. Figure~\ref{fig-sca} shows also that the quantity  $(\eta^2/{\cal -L})^{1/2}$ reaches a region where it is approximately linear in $t$. 
The two proportionality constants,  $\gamma_m$  and  $\gamma_{\cal L}$, refer to two different quantities in the simulations: the former refers to the number of monopoles in the network while the second refers to the typical  intermonopole distance. The values  of $\gamma_m$ and $\gamma_{\cal L}$ for each different type of simulation are shown in table~\ref{tab-gam}.

We achieve scaling for all different cases simulated, though at different stages. The cases with $s=0$ are the ones which reach scaling fastest, and therefore, have the longest dynamical range simulated. For the $s=1$ case, the radiation simulations achieve scaling earlier than the matter case. Besides, in  the matter $s=1$ case, the lagrangian measure of scaling has much more structure than the other cases, i.e., there are some small spikes along the straight line.  As an example, we plot the scaling regime for the two extreme cases in figure~\ref{fig-sca}:  radiation era with $s=0$, which reaches scaling fastest and the scaling line is rather smooth,  and the Matter era  with $s=1$, which has the latest onset of a bumpier looking scaling. In the other  two  cases the scaling regime is as smooth as in radiation era with $s=0$ but in the case of radiation era with $s=1$, scaling is reached later.

Note that  the $s=1$ case is the closest to the real equations of motion, but the dynamical range obtained is shorter than that of the $s=0$ case. Therefore,  we have more 'realistic' data with $s=1$, though with shorter dynamical range and less statistics; and data obtained from  modified equations of motion with $s=0$ (and therefore with a higher level of modelling), but with higher dynamical range and more statistics. We will show in the following sections that both approaches give compatible results. This can also be seen in table \ref{tab-gam}, where the values  of $\gamma_m$ and $\gamma_{\cal L}$ for each different types of simulation are shown to be compatible within the same cosmological era. 

\begin{table*}
\begin{center}
\scalebox{0.8}{
\begin{tabular}{ |c|c|c|c|c|}\hline
 &  \multicolumn{2}{|c|}{$\gamma_m$} & \multicolumn{2}{|c|}{$\gamma_{\cal L}$}\\ \cline{2-5}
s &   Radiation & Matter& Radiation & Matter \\ \hline
0 & 0.72 $\pm$ 0.06 & 0.65 $\pm$ 0.04 & 0.1$\pm$ 0.05&  0.20 $\pm$ 0.05\\ \hline
1 &  0.76 $\pm$ 0.03 & 0.58 $\pm$ 0.03 & 0.1$\pm$0.05 &  0.25 $\pm$0.05\\ \hline
\end{tabular}
\caption{\label{tab-gam} Values of   $\gamma_m$ and $\gamma_{\cal L}$ for matter and radiation eras. Note that for a given era, the values obtained for different $s$ are compatible. }
}
\end{center}
\end{table*}

Once the system reaches scaling, quantities of interest can be measured: for example, monopole velocities. There are several systematic errors that the reader should be aware of. On the one hand, there are numerical errors inherent to the simulation of the dynamics of the system. By these we mean errors arising form the discretisation of the equations, errors due to the limited dynamical range and errors coming from the Press-Ryden-Spergel algorithm.  On the other, there will be errors coming from the procedure used in tracking each monopole's trajectory, as explained in the next section.

\section{Monopole velocity estimators}
\label{tv}

The magnitude of interest in this work is the averaged network velocity of the monopoles. There are two procedures  proposed in the literature to obtain the network velocity, which we will revisit momentarily. But first, we will describe in detail the novel procedure proposed in this work, called the {\it Monopole--Tracking} Method.

\subsection{Monopole--Tracking Velocity }

In this method we calculate the monopole velocity by measuring directly  the distance traveled by a monopole in a specific period of time. In order to do so, we need to pinpoint where each monopole is at every time step,  we need to determine where it moves to at the following  step and we need to track all the steps of each monopole.

Note that  we are evolving  field values in each lattice point, and not point-like global monopoles. For this reason every time step we have to identify monopole positions within the lattice; that is, we have to translate the information from field values to monopole positions. This translation can be done by working out the topological charge  in each one of the lattice points. The topological charge of monopoles is given by the surface integral \cite{Vilenkin:2000jqa}
\be
N=\frac{1}{8\pi}\oint dS^{ij}|\Phi|^{-3}\epsilon_{abc}\Phi^a\partial_i\Phi^b\partial_j\Phi^c.
\label{tp}
\ee
Thus, in order to determine the location of a monopole, we calculate whether the integral (\ref{tp}) has a non-zero value. 
Actually, since we are in a discretized environment,  we do not use directly  formula (\ref{tp}), but rather a discrete version of it (see Appendix B in \cite{Antunes:2002ss}). We therefore are computing the topological charge  inside a cube of dimensions $dx^3$;  cubes that are characterised by their vertices. 
 
Using this method we are able to detect   positions in the lattice with non-zero topological charge; i.e., we detect the positions of both monopoles (positive topological charge) as well as anti-monopoles (negative topological charge).  As mentioned before,     the monopole number ${\cal N}$   will be given by \be
{\cal N}= m+\bar{m},
\label{N}
\ee
where $m$ refers to the number of monopoles and $\bar{m}$ to the number of anti-monopoles in the simulation. The left pane of fig.~\ref{v-ko} shows a snapshot of a simulation where monopoles and antimonopoles have been detected.

The detection of monopoles is performed during run-time, together with the evolution of the equations of motion. In principle one would like to detect monopoles every time step. However, the topological charge calculation is very time consuming, and instead of computing it for every time step, we only computed it for every $t_{s}$ time interval. Table~\ref{params} shows the values of $t_s$ for every case. At early times a finer $t_s$ is necessary because the density of monopoles is higher, and at later times a coarser search is enough.  We have checked in a few simulations  that the results obtained using these $t_s$ gives the same result as performing it at every single time step.

Once all monopoles in all time steps have been detected, we start the reconstruction of the world line of each monopole. In order to do so, we need to identify where each monopole has travelled from time $t_1$  to time $t_2$; or in other words, we have to identify all monopoles at time $t_1$ with all monopoles at time $t_2$. This matching procedure is achieved by computing the distances between monopoles at time $t_1$ and those at time $t_2$; and pairing the ones that are closest to each other. We are  aided by these two considerations: first, we check that the topological charge of the two monopoles to be paired is the same, that is, a monopole does not turn into an antimonopole as it moves, or viceversa. The other consideration is that of maximum distance: we do not want a monopole to travel much faster than the speed of light.

This second consideration needs some explanation. Imagine the following situation: a monopole just met an antimonopole in the simulation just after time $t_1$. Thus,  in the next time step  to be analyzed ($t_2$) they have both annihilated  each other. However, our procedure is unaware of the annihilation and would still find the monopole at time $t_2$ that is closest to the monopole at $t_1$ under consideration. Since that monopole is nowhere to be found in $t_2$, it will pair it with another monopole, with the one that  is the closest, even though that would be another monopole. If we did not account for this kind of event, we would obtain extremely high velocities when annihilation events happen. In order to avoid these, we suppress all monopole pairings that  mean velocities higher than 1.5 times the speed of light\footnote{We also checked the results with a cutoff of $c$, $2c$ and $3c$. We found out that a cutoff of $c$ gave the same results as $1.5c$, but decided to keep the higher cutoff to show explicitly that we were allowing for superluminal velocities. In the higher cutoff cases ($2c$ and $3c$)  the velocities obtained were higher, but in all cases it was due to  monopoles being matched to the wrong monopole; not because their actual velocity was superluminal.}. We allow for speeds higher than the speed of light since previous works \cite{Yamaguchi:2001xn} measured velocities compatible with superluminal velocities. 

Once the monopoles (and antimonopoles) have been paired, we then start reconstructing their path. First we match all monopoles at time $t_1$ with those at time $t_2$; then we match those at time $t_2$ with those at time $t_3$; and so forth. This way we can obtain the whole worldline of each  monopole in the simulation box. 

There is, however, another subtlety: since we are in a discretized situation, there are instances when a monopole is not inside a $dx^3$ cube, but it is moving through the face that divides two $dx^3$ cubes. In those cases we can fail to detect a monopole. In those instances, the corresponding monopole at a previous time-step remains unmatched,  and the monopole at the following time-step may seem to have appeared from thin air.  In order to account for this problem, whenever a monopole failed to be matched with a monopole from the {\it previous}  step, we looked at two time steps back to match it with a monopole there.  With this method we observe that all instances of monopoles coming out of nowhere were solved.

Once we have the paths of all the monopoles in the simulation box, the velocities can be obtained straightforwardly by dividing the distance traveled by the monopole by the time lapsed to travel that distance.  Then we can average over all the velocities of the monopoles to obtain a network average velocity. This number can be used for the calibration of the effective model of global monopoles; and it can also be used to compare   with the velocities obtained with the other network velocity estimators explained below.

In order to compare the results obtained by our novel method with those already in the literature, we  average the velocity of each monopole to obtain the average velocity of the network. Actually, the calibration of the effective model for global monopoles needs this one average network velocity number.

\subsection{Local Velocity Estimator\label{ya-ve}}

This method  was proposed by  Yamaguchi \cite{Yamaguchi:2001xn} to estimate global monopole network velocities using field values.  This method relies on the fact that all the information about the global monopoles is included in the  scalar fields. We summarize the method in the following lines and direct the interested reader to  \cite{Yamaguchi:2001xn} for details; and then explain an improvement on the original method.

Let us assume that the monopole at $t_0$ is located at ${\bf x_0}$, and at a sufficiently close time $t$ is at ${\bf x}$. Let us  
 expand the scalar fields $\phi^i({\bf x}, t)$ around $\phi^i({\bf x_0}, t_0)$ up to first order, 
\begin{equation}
\phi^i({\bf x}, t) \simeq \phi^i({\bf x_0}, t_0)+\nabla  \phi^i({\bf x_0}, t_0) ({\bf x}-{\bf x_0}) + \dot{\phi}^i({\bf x_0}, t_0) (t-t_0)\,.
\end{equation}
Bearing in mind that the scalar fields vanish at the position of the monopole, we have that at times $t_0$ and $t$ both  $\phi^i({\bf x_0}, t_0)=0$ and  $\phi^i({\bf x}, t)=0$, which lead us to 
\begin{equation}
\nabla  \phi^i({\bf x_0}, t_0) ({\bf x}-{\bf x_0}) + \dot{\phi}^i({\bf x_0}, t_0) (t-t_0)=0\,.
\end{equation}

Solving those linear equations in terms of the field values $\nabla  \phi^i({\bf x_0}, t_0)$ and $ \dot{\phi}^i({\bf x_0}, t_0)$, the velocity of global monopoles can  be estimated as
\begin{equation}
v=\frac{|{\bf x}-{\bf x_0}|}{t-t_0}\,.
\end{equation}

As explained in \cite{Yamaguchi:2001xn}, one main source of errors for the Local Velocity Estimator (LVE) comes from the fact that a monopole does not generally lie just on the lattice point in a simulation, but the actual zero of the fields is between lattice points, and therefore the LVE approximation is not accurate. Thus, the monopole detection algorithm used in \cite{Yamaguchi:2001xn} was prone to errors, and at some lattice points the velocities obtained  were extraordinarily large, even at points that had nothing to do with a monopole core  \cite{Yamaguchi:2001xn}. 

In order to reduce these errors we have refined Yamaguchi's approach. Since we do have the information (at run-time) of the location of the monopoles via their topological charge, we only compute velocities using LVE at the eight vertices of the cubic cell where the monopole is located. We then average the velocity over the eight points. This improves the prediction, but  still the error in the velocity of each individual monopole  is quite significant, and we still get situations with velocities that are very high. We then disregard instance where  the value is greater than 1.5  and  average over all monopoles to obtain a meaningful estimate. Thus, we will report only the average velocity of the monopole network.

\subsection{Average Velocity Field Estimator\label{mA}}

This method was originaly proposed by \cite{Hindmarsh:2008dw} for Abelian Higgs cosmic strings,  but   it can readily be used for global monopoles \cite{HPC}.  As LVE, the Average Velocity Field Estimator (AVFE) uses the values of the fields at each lattice point to estimate network velocities. 

In the AVFE, the  velocity is obtained directly by the local values of the derivatives of the fields.  We will use  the momentum ${\bf \Pi}=\dot{\bf \Phi}$ and the spatial gradients of the fields  $ {\bf \partial \Phi}=\frac{\partial {\bf \Phi}}{\partial {\bf x}}$ to estimate velocities. Since we are interested in the velocities of the monopoles, we want our estimator to pinpoint regions with monopoles. In order to do so, we weight the derivates  of the fields with the potential energy, since the potential energy peaks at sites where the fields are close to zero, denoting (mostly) the core of the monopole. We denote the  weighting  of a field ${\cal X}$ with respect of  the potential energy ${\cal V}$  as
\begin{equation}
{\cal X}_{\cal V}=\frac{\int d^3 x \; {\cal X} \;{\cal V}}{\int d^3x \; {\cal V}}\,.
\label{we}
\end{equation}
Thus,  by defining the ratio $R_{\cal V}={\bf \Pi}_{\cal V}^2/{\bf \partial \Phi}^2_{\cal V}$ , we can obtain our AVFE   $\langle \dot{\bf x}^2 \rangle$  as:
\begin{equation}
 \langle \dot{\bf x}^2 \rangle= \frac{3R_{\cal V}}{1+2R_{\cal V}}\,.
\label{vA}
\end{equation}

Monopoles can be understood as  concentrations of potential energy, that is, regions where the potential is out of its minimum. Therefore,  the potential  is a good weighting field, which ensures that the points on the lattice containing monopoles are selected, and regions without monopoles are rejected. 

\begin{figure}[!htb]
\includegraphics[width=0.46\textwidth]{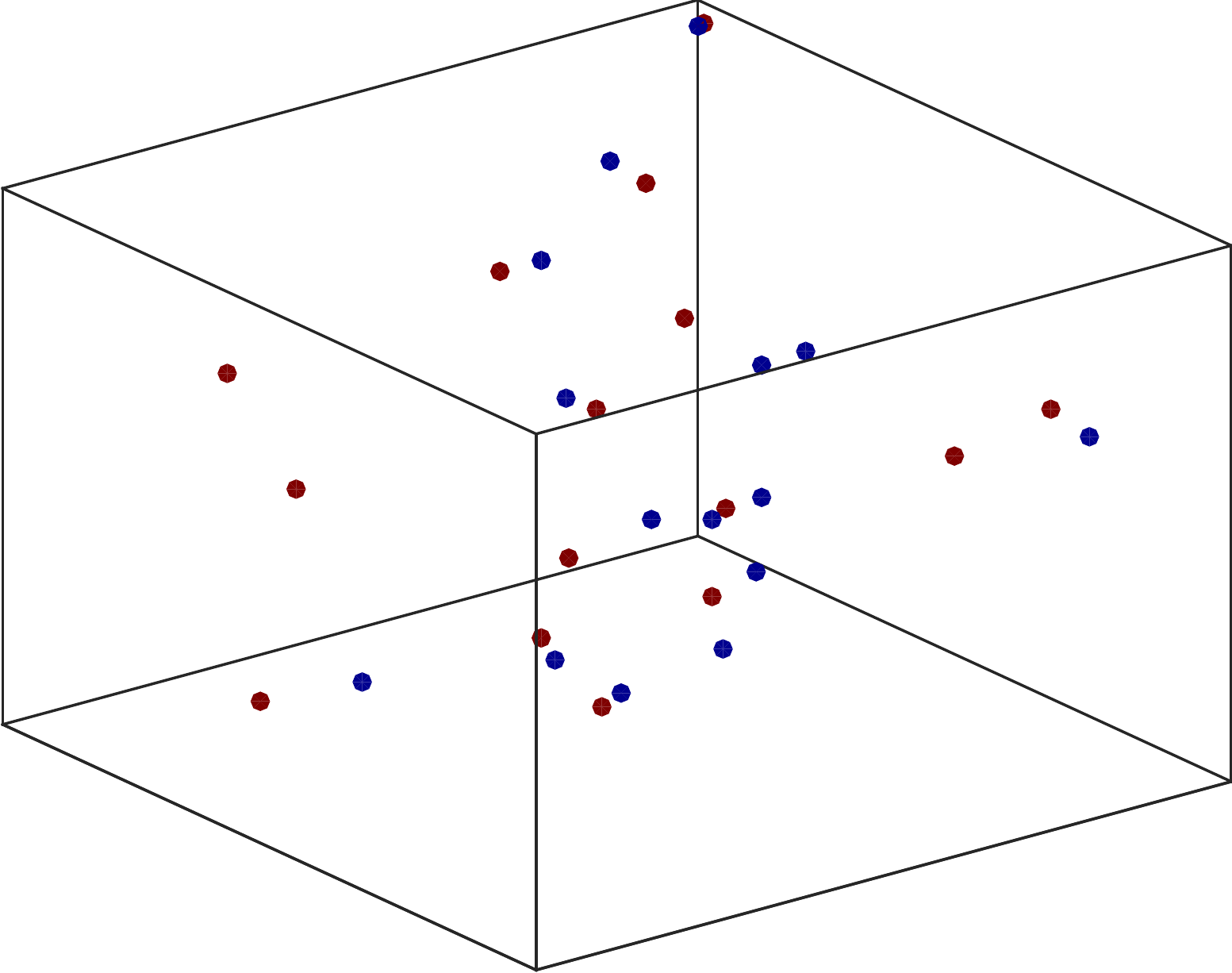}
\includegraphics[width=0.46\textwidth]{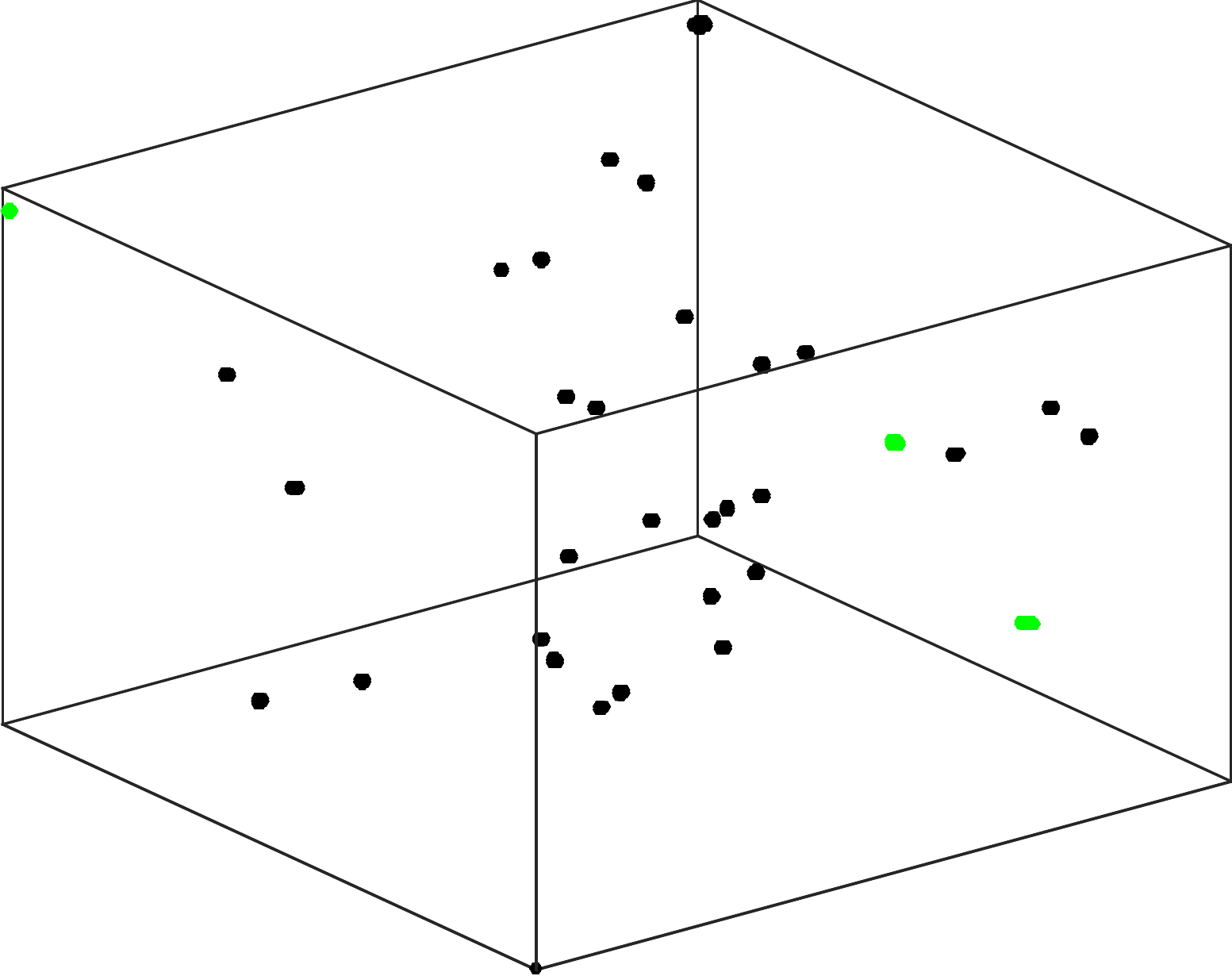}
\caption{In the left pane we show the simulation box after working out the topological charge in each point of the box:  red dots represent antimonopoles (points where the topological charge is -1), and  blue dots represent monopoles (points where the topological charge is 1). The right pane show the same simulation at the same time, but in this case we show regions where the value of the potential energy is high. It is clear that all the monopoles and antimonopoles in the left pane have been recovered (black dots). There are few regions (in green) that do not correspond to a monopole detected by measuring the topological charge. These can be regions where a monopole-antimonopole pair has just annihilated, or monopoles that are crossing the face of a lattice cell, and even if they are not detected by the topological charge, they are detected by the potential energy. \label{v-ko} }
\end{figure}

We can obtain a visual confirmation that this  indeed succeeds in selecting the regions where there are monopoles.  In figure~\ref{v-ko} we plot a configuration of monopoles as detected by calculating topological charge, and also we plot the values of the potential energy that are greater than a threshold.  It is clear that   the potential energy  mimics the monopole positions satisfactorily.  We observe, however, that even though all the monopoles are recovered, there are some regions that do not correspond to a monopole (in green, in the figure). Those points would correspond to, for example, a monopole-antimonopole pair that have annihilated, and there is still considerable potential energy in the region, even though there is no topological charge. It could also be the case that the monopole is crossing a face of a cube, and thus we do not detect it by the topological charge, but the potential energy is able to pinpoint it. In any case,  we will see in Section \ref{re-v}  that the effect of these regions  is very small.

\section{Velocity Results \label{re-v}}
We have used the previously described three methods to estimate the velocity of the network of monopoles. Two of the methods, the {\it Local  Velocity Estimator}  and the {\it Average Velocity Field Estimator} are able to give us a network average velocity only; whereas  our new estimator, the {\it Monopole--Tracking Velocity} estimator, is able to obtain velocities of single monopoles, which can be obviously averaged over to obtain a network velocity estimator.

\begin{figure}[!htb]
\includegraphics[width=0.45\textwidth]{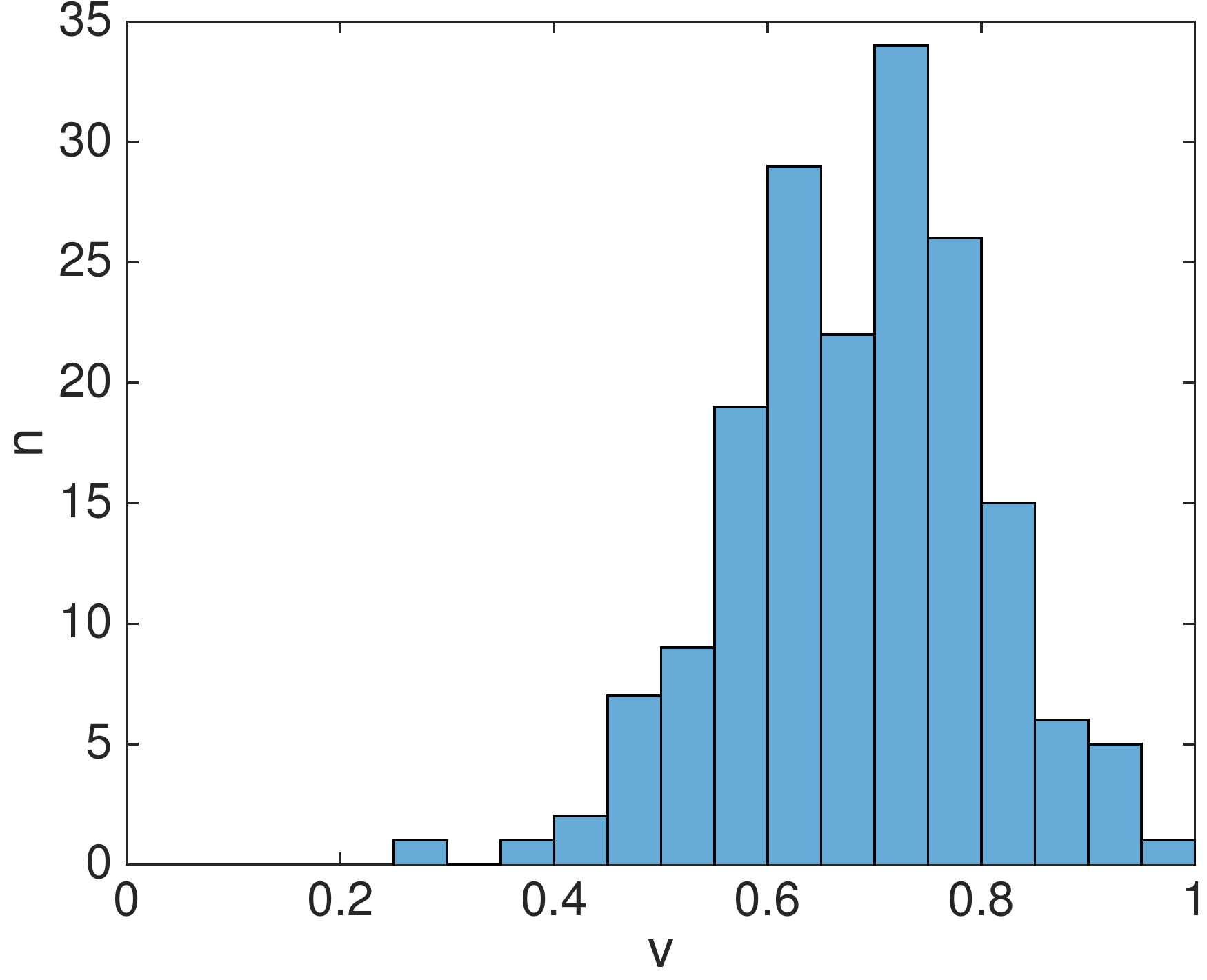}\hspace{0.5cm}
\includegraphics[width=0.45\textwidth]{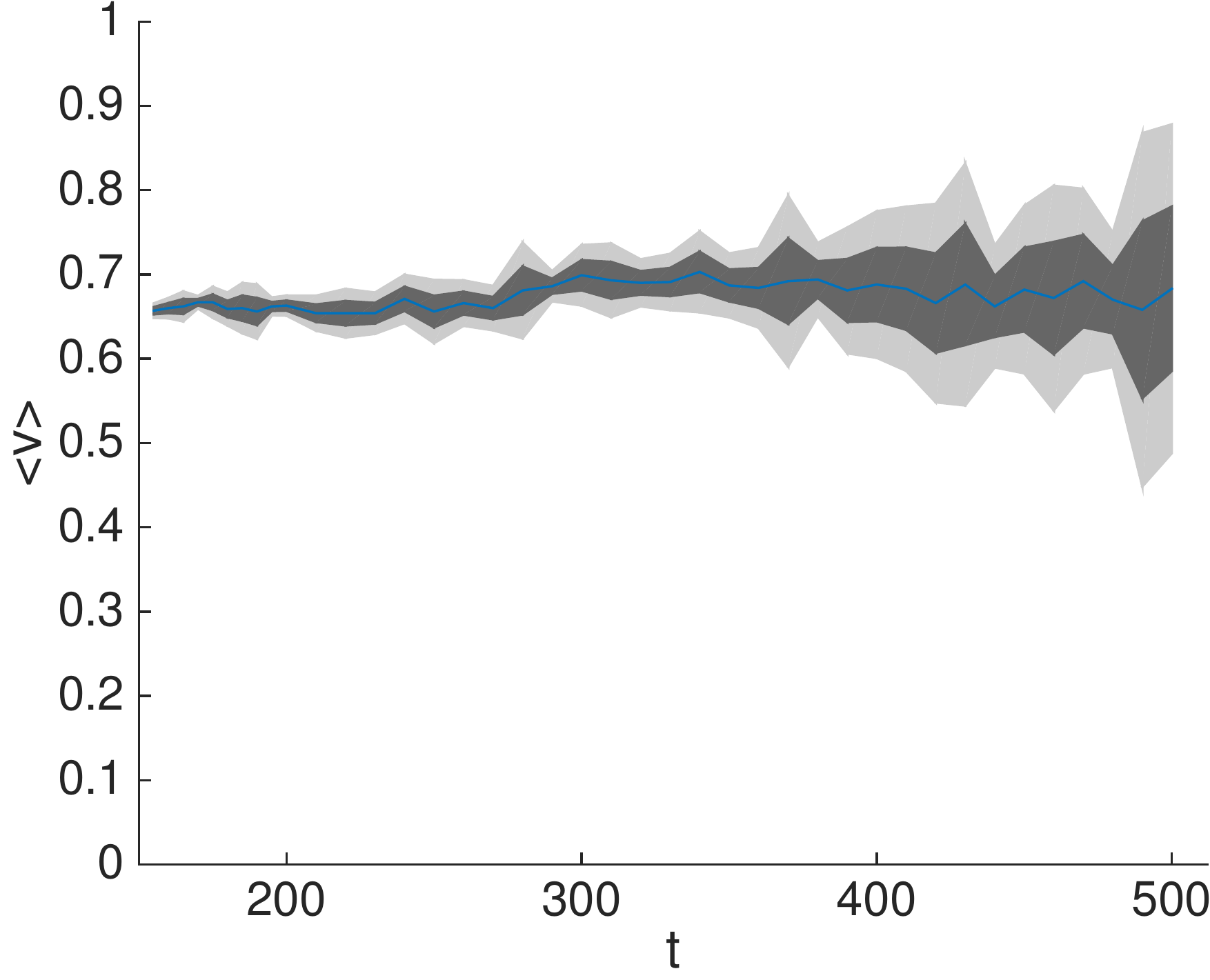}
\caption{Values of the velocities for radiation era with $s=0$. In the left plot  the number of monopoles is divided in velocity intervals:  the intervals are  20 equally spaced  bins in the  $[0,1]$ range. These velocities are obtained by dividing the total length travelled by the time it takes. In the right plot we show the average of the instantaneous velocity over all monopoles in each time step.  Note that even though we allow for monopoles to have velocities up to $1.5c$, we have not observed any monopole with $v>1$ in any of the simulations we have performed. \label{fig-v} }
\end{figure}

We used Monopole--Tracking Velocity  to compute the velocities of each one of the monopoles in the simulation box for every time step. This information can be presented in two different manners.   In fig.~\ref{fig-v}  we have depicted the case for radiation era and $s=0$ as an example; the behaviour is similar for the other case. In the left panel we consider the  average  velocity of each monopole, and plot a histogram where the bins are 20 equally spaced bins in the $[0,1]$ range. To obtain the velocity of each monopole, we track the path  of each monopole and divide the total length of the path of each monopole by the time it took to cover that distance.
In the right panel, we average over the instantaneous velocity of all the monopoles for every time step, i.e., at every time-step we calculate the velocity of each monopole, and take the average of all monopoles in the box. 

The values of the velocities obtained for every case can be found in Table~\ref{tab-vel1}. We have included the velocities obtained by a) the average velocity of each monopole, as obtained by dividing the total length travelled by the time taking to cover it (Total-length) and b) the average of all instantaneous velocities (Instantaneous). We also take two types of averages: a)   averaging over the absolute value of the velocities ($\sum_{i=0}^{\cal N} |v_i|/{\cal N}$), and b) using the root mean square (RMS) of the ensemble ($\sqrt{\sum_{i=0}^{\cal N} v_i^2/{\cal N}}$).  

Our results show that the global monopole velocity is sub-luminal; not a single monopole velocity (in any of the cases studied) was measured to be higher than 1. The average velocity is consistent with a constant velocity, and that velocity coincides with the mode of the velocity distribution (see the histogram in Fig.~\ref{fig-v}). The values of the velocities obtained for every case can be found in Table~\ref{tab-vel1}, both averaging over the absolute value of the velocities, and also by the RMS velocity of the ensemble.  

The errors quoted  include statistical errors as well as   an estimation of the systematic errors. These systematic errors come mainly from the identification method: we are assuming that the monopole is at a specific point in the lattice, whereas in reality it can be anywhere inside the $dx^3$ cube specified by that lattice point; therefore, there is an error in the length of the path due to the discretization of the lattice.  Another possible ingredient that might be included  into the errors is the correlation between the measurement of   individual monopole velocities. Monopoles are interacting with each other, and it is not surprising to think that there might be some connection between the errors; but we expect this to not change drastically the error budget.

\begin{table*}[!htb]
\begin{center}
\begin{tabular}{ |l| c| c|c|c| }
\cline{2-5}

  \multicolumn{1}{c|}{}  & \multicolumn{2}{|c|}{Radiation}  & \multicolumn{2}{|c|}{Matter}   \\  \cline{2-5}
    \multicolumn{1}{c|}{}  & $s=0$ & $s=1$ & $s=0$ & $s=1$  \\  \cline{2-5}
     \multicolumn{1}{c|}{}  & \multicolumn{4}{|c|}{Average of absolute velocities}   \\\hline
Total-length  & $0.70\pm 0.05$&  $0.70  \pm 0.05$ &  $0.62 \pm 0.05$ & $ {0.55  \pm 0.05 }$  \\ \hline
Instantaneous & $0.70 \pm 0.09$&  $0.70 \pm 0.09$ &  $0.62 \pm 0.09$ & ${0.55 \pm 0.09 }$   \\ \hline
\multicolumn{1}{c|}{}  & \multicolumn{4}{|c|}{Root mean square velocities}   \\  \hline
Total-length & $0.71\pm 0.07$&  $0.70 \pm 0.09$ &  $0.63 \pm 0.07$ & $ {0.55  \pm 0.09 }$  \\ \hline
Instantaneous & $0.72 \pm0.07$ &  $0.72 \pm 0.07$  &  $0.65 \pm 0.07$ & ${0.59 \pm 0.06 }$   \\ \hline
Total & $0.71\pm0.05$ & $0.72\pm0.06$ & $0.64\pm0.05$& $0.57\pm0.05$\\\hline
\end{tabular}
\caption{\label{tab-vel1} Values of the velocities for matter and radiation eras, and  for $s=0$ and $s=1$ cases, using the  Monopole--Tracking method (Track) estimator. We report velocities obtained by considering the average velocities of the monopoles obtained by the time spent in travelling the length of their path  (total-length) and by the average of the instantaneous velocities of the monopoles in each time-step (Instantaneous). We also report the average of the absolute value of the velocities, and RMS velocity value. We can see that the values given by the different methods are compatible with each other, including the values for $s=0$ and $s=1$. The last row shows the combined RMS velocities, combining both averaging and instantaneous velocities. Typical velocities during matter domination are somewhat slower, consistent with the higher Hubble damping.}
\end{center}
\end{table*}

\begin{figure}[!htb]
\includegraphics[width=7.5cm]{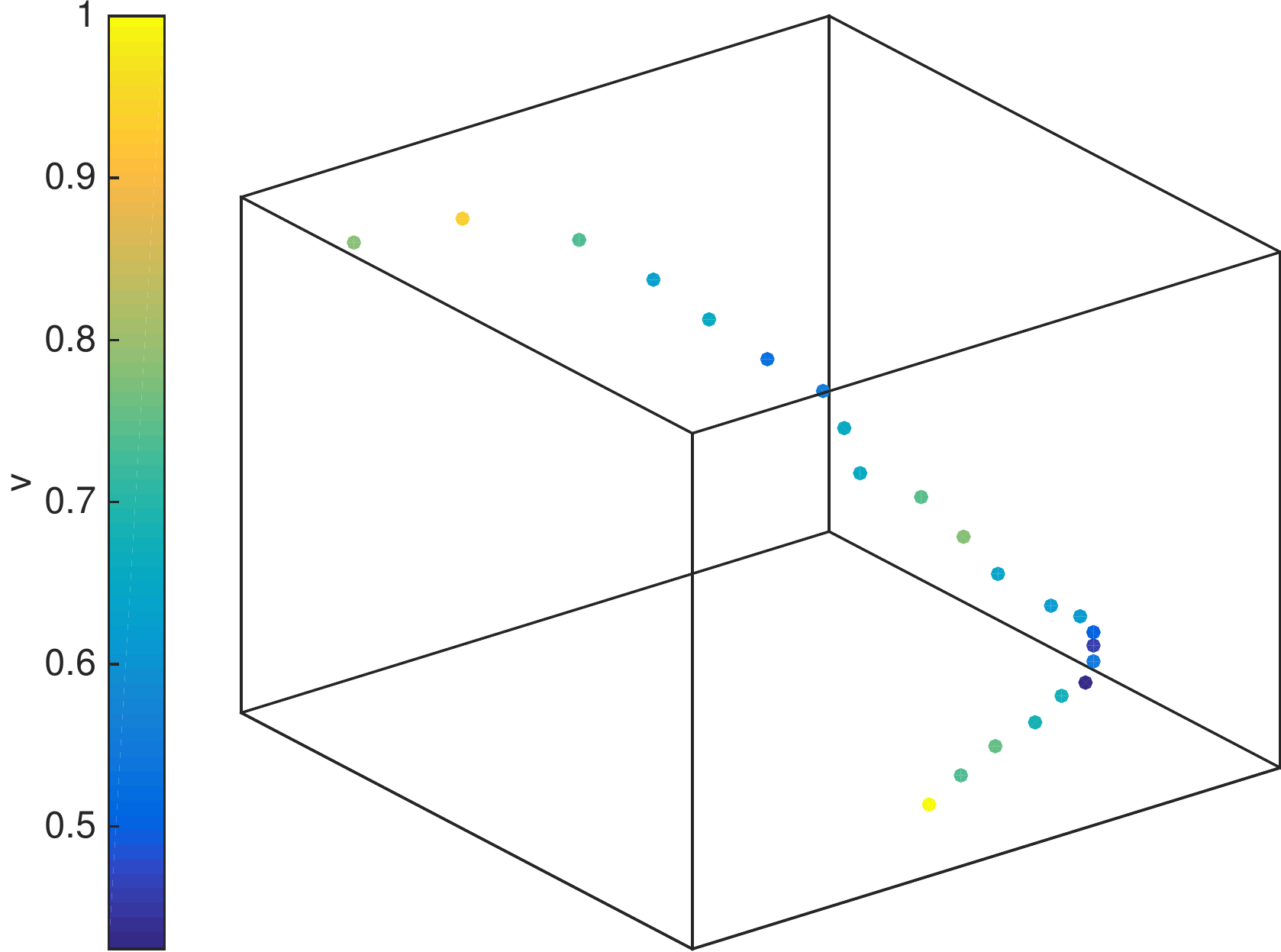}\hspace{0.5cm}
\includegraphics[width=7cm]{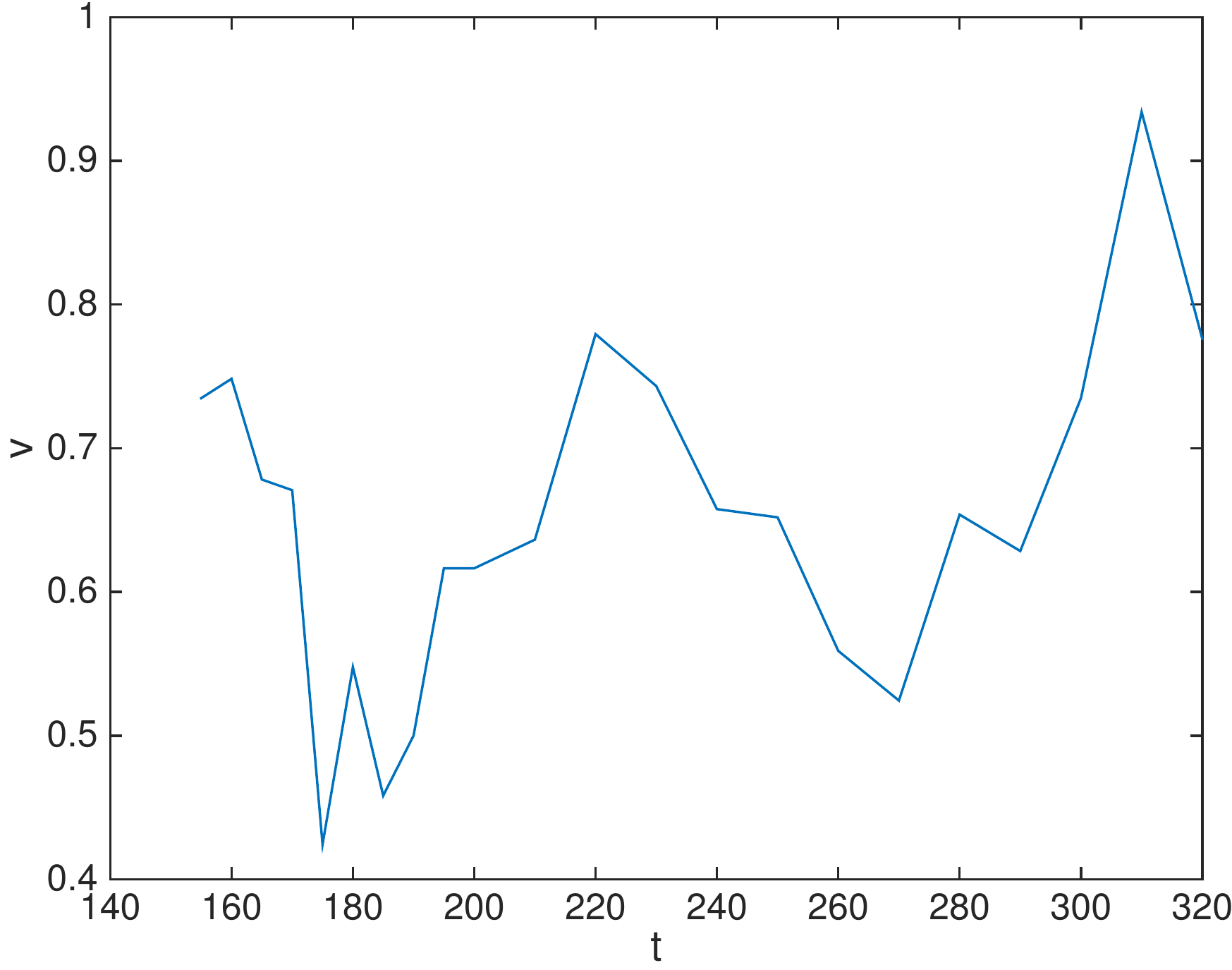}
\caption{\label{path-v}The left pane shows the path described by a monopole during its evolution. The initial time step is located at the bottom of the box, so the path  travelled by the monopole has to be understood to go from bottom to top. The color indicates the velocity of the monopole at each time step, which can be better viewed in the right pane,  showing the non-trivial velocity history of the monopole.}
\label{fig-path}
\end{figure}

The monopole--tracking method also allows to study the velocity history of individual  monopoles, showing that the situation is much richer than a single network average can show. For example,  Fig.~\ref{fig-path} shows the path of a typical monopole, with the value of its velocity in each time step. It can be seen that the velocity history of the monopole is highly non-trivial.  The monopole travels in a more-or-less straight line, then reduces its velocity to change directions, and then continues in another straight line. This can be understood by considering that the monopole may be travelling to meet an antimonopole, but before it gets there, the antimonopole has annihilated with another monopole. The field configuration around the first monopole then reorganises and the monopole heads towards another antimonopole.

We also used  our improved version of the Local Velocity estimator  (LVE) and the Average Velocity Field Estimator (AVFE) to obtain an average network velocity. Table \ref{tab-vel} shows the values of velocities obtained with all
different methods and configurations. The errors for the LVE and AVFE methods are just statistical errors. As in the previous case we are assuming that the measurements are independent, though clearly there may be some correlation between them.

\begin{table*}[!htb]
\begin{center}
\begin{tabular}{ |l| c  | c|c|c|c|c|}
\cline{2-7}
 \multicolumn{1}{c|}{}  & \multicolumn{3}{|c|}{s=0}  & \multicolumn{3}{|c|}{s=1}\\  \cline{2-7}
\multicolumn{1}{c|}{} & Track &  LVE & AVFE &  Track  & LVE &AVFE\\ \hline
 Radiation  & $0.71\pm0.05$&  $0.7 \pm 0.1$ &  $0.85 \pm 0.06$ & $ 0.72 \pm 0.06 $ &   $ 0.7 \pm 0.1 $  &  $ {0.90 \pm 0.09}$ \\ \hline
Matter & $0.64 \pm0.05$ &  $0.6 \pm0.1$  &  $0.76 \pm 0.08$ & ${0.57 \pm 0.05 }$  & ${0.6 \pm 0.1 }$ & ${0.80 \pm 0.09 }$\\ \hline
\end{tabular}
\caption{\label{tab-vel} Values of the RMS velocities for matter and radiation eras using the methods described in the text: the Monopole--Tracking method (Track), the Local Velocity estimator (LVE) and the Average Velocity Field estimator (AVFE). The monopole--tracking values quoted here are the average of the RMS velocities obtained by a combination of 'average' and 'instantaneous' averages (see Table~\ref{tab-vel1}). We can see that the values given by the different methods are compatible with each other, including the values for $s=0$ and $s=1$.}
\end{center}
\end{table*}

We see  that the results obtained using the three different methods are compatible with each other (within 1$-\sigma$ for $s=0$ and 2$-\sigma$ for $s=1$); and in all cases the velocities are subluminal. We also see that the velocity of monopoles in the radiation  era is higher than in the matter era, consistent with the lower Hubble friction in radiated--dominated expansion. 

The agreement between the Tracking and the LVE methods is remarkable.
 The Average Velocity Field Estimator, however, tends to overestimate the velocity of the monopoles, but this is not surprising since the weighting method encapsulates points in the simulation which have some velocity but are not monopoles. For example, the AVFE includes regions where a  monopole-antimonopole  have annihilated and have left some temporary residual 
potential energy, before it decays away like radiation. It is reasonable to think that those regions would contribute with high velocity to the average.
 
We can also see that the results obtained using the $s=0$  and $s=1$ cases are very similar. Bear in mind the dilemma in deciding what procedure to use: either we simulated for a longer dynamical range with modified equations of motion; or we simulated the equations of motion but for a shorter dynamical range. Our results show that both methods render results that are compatible with each other. Therefore, it seems appropriate to take advantage of both the larger dynamical range and the fact that the true equations of motion can be simulated to report the velocity of the global monopoles combining all simulations together. 

The average velocities of a network of global monopoles, combining the $s=0$ and $s=1$ simulations, together with the three velocity estimation methods described above, for radiation and matter dominated epochs, respectively, are   
\begin{equation}
v_r=0.76\pm0.07\,,\qquad v_m=0.65\pm0.08\,.
\label{vels}
\end{equation}

The errors are obtained by averaging over all the data (three methods, five simulations and both values of $s$). We have been conservative  in the errors quoted, bearing in mind that the measurements may not be totally uncorrelated, and that there may be some (small) changes due to correlations. 

\section{Calibration of the VOS model for global monopoles\label{an-ca}}

Analytic models are effective models that capture the properties of a network of defects into evolution equations for macroscopic quantities that describe the network. In some sense, analytic models concentrate on the thermodynamics of the network instead of on the details of the dynamics.  The evolution equations  have some phenomenological parameters to be determined by the true microphysics of the system. Our simulations enable us to determine those phenomenological parameters.

The analytic model for global monopoles was presented in \cite{Martins:2008zz}. This method is based on the velocity dependent one-scale model of Martins and Shellard \cite{Martins:1996jp,Martins:2000cs}, which recognises that in order to be able to quantitatively describe the whole cosmology history of the networks one must be able to describe the evolution of the defects velocities. To do so, it retains Kibble's \cite{Kibble:1984hp} assumptions on the existence of a single length scale but adds the RMS velocity as a second macroscopic quantity. In this model the length scale, $L$, and the RMS velocity, $v$, are described by the following equations  
\be
3\frac{dL}{d\tau}=3HL+v^2\frac{L}{l_d}+cv,
\label{eq-l}
\ee
\be
\frac{dv}{d\tau}=(1-v^2)\Big[\frac{k}{L}\Big( \frac{L}{d_H}\Big)^{3/2}- \frac{v}{l_d}\Big].
\label{eq-v}
\ee
The parameters $H$ and $d_H$ are the Hubble parameter and the Hubble horizon size, and $\tau$ is the physical time (as opposed to the conformal time in Eq.~(\ref{eom})). The overall damping length which includes both the effect of Hubble damping and of friction due to particle scattering is parametrised by $l_d$. Finally, $c$ and $k$ are the parameters governing the phenomenological terms, and these are the  parameters that we will calibrate using  our data.  

The term involving $c$ is associated with energy losses from monopole-antimonopole annihilation; in some sense, it depends on short distance physics. The term involving $k$ is the acceleration due to the forces among monopoles. These forces are approximately independent of distance but, because global monopoles have linearly divergent gradient energy, the "mass" at a given scale $L$ grows linearly with $L$. The resulting acceleration term is of the form $\sim k/L$, and is corrected by a $1 / \sqrt {\cal N}$ factor to account for the combined effect of interactions with multiple  monopoles. In ref. \cite{Martins:2008zz} the authors considered how sensitive the solutions are  to  the modelling of the parameters, and found that  the final characterization of the network is much more dependent on details of the term involving  $k$ than those of the term involving $c$. 

We are interested in identifying the scaling solutions for this model.  In order to do so, we first observe that in our simulation  the  characteristic length of the network (see Eq.~\ref{gammat})  is proportional to time\footnote{Note that the parameter $\epsilon$ is the analogous to the parameter $\gamma_t$ in  Eq.~\ref{gammat}, but time is now physical instead of conformal.} and that the  velocities are constant:
\be
L=\epsilon \tau\,, \qquad v=v_0=const.
\ee

As shown in \cite{Martins:2008zz}, for expansion rates of the form $a(\tau) \propto \tau^\lambda$ with $\lambda < 3/4$ (which includes matter-- and radiation--domination), the equations  (\ref{eq-l}) -  (\ref{eq-v})  admit two branches of scaling solutions:  an ultrarrelativistic one with $v_0 = 1$ and a subluminal one with $v_0 <1$. We will not study the values of the parameters for the luminal branch, since our simulations show no evidence of its existence.  For the subluminal branch we can read off the values of $c$ and $k$,
\be
c=\frac{\epsilon v_0}{3(1-\lambda)-\lambda v_0^2}\,,\qquad k=\frac{\lambda v_0}{(1-\lambda)^{3/2}\epsilon^{1/2}}.
\label{eq-pa}
\ee

\begin{table*}[!htb]
\begin{center}
\scalebox{0.8}{
\begin{tabular}{ |l| c  | c|c|c|c|c|}
\cline{2-7}
 \multicolumn{1}{c|}{}  & \multicolumn{3}{|c|}{s=0}  & \multicolumn{3}{|c|}{s=1}\\  \cline{2-7}
\multicolumn{1}{c|}{} & $\epsilon$ & c &k & $\epsilon$ & c & k \\ \hline
 Radiation  & 1.42$\pm$ 0.09& 0.80$\pm$0.06 & 0.76 $\pm$ 0.02 & 1.53 $\pm$ 0.04& 0.93$\pm$0.04 & 0.92 $\pm$ 0.02 \\ \hline
Matter & 1.97 $\pm$0.09 & 0.71$\pm$ 0.03 & 1.55 $\pm$ 0.04 & 2.00 $\pm$0.06 & 0.65$\pm$ 0.02 & 1.42$\pm$ 0.02 \\ \hline
\end{tabular}
\caption{\label{tab-an} Values of the analytic parameters for matter ($\lambda = 2/3$) and radiation ($\lambda = 1/2$) and for $s=0$ and $s=1$.
}}
\end{center}
\end{table*}

In Table \ref{tab-an} we show the results obtained using equations (\ref{eq-pa}) and our numerical results. To obtain the values for $\epsilon$ we have used the slopes of the scaling curves, as given in Table~\ref{tab-gam}.  As for the values for the velocity, we have used  the velocity obtained by averaging over all three methods of RMS velocity estimation discussed in section~\ref{re-v} and  shown in Table~\ref{tab-vel}.

Since the velocity results for both $s=0$ and $s=1$ are compatible with each other, and they  complement each other, we could also try to combine all simulations with different $s$ to obtain an estimate of parameters $c$ and $k$. This way we obtain a more conservative value of the errors on the parameters. Averaging over all simulations, both with $s=0$ and $s=1$, we obtain the values shown in table~\ref{tab-eps}.
\begin{table*}[h]
\begin{center}
\begin{tabular}{ |l| c  | c|c|c|}
\cline{2-4}
\multicolumn{1}{c|}{} & $\epsilon$ & c &k  \\ \hline
 Radiation  & 1.47 $\pm$ 0.09& 0.9 $\pm$ 0.2 & 0.9 $\pm$ 0.1  \\ \hline
Matter & 1.98 $\pm$ 0.07 & 0.7 $\pm$ 0.2 & 1.6 $\pm$ 0.2 \\ \hline
\end{tabular}
\caption{\label{tab-eps} Values of the analytic parameters for matter  and radiation averaging over all simulations with $s=0$ and $s=1$.  We first average over all velocities, and then use that average (with errors)  to obtain the value of $c$ and $k$.}
\end{center}
\end{table*}

We see that the values of $c$ in radiation and matter are compatible with each other, but even when treating the errors conservatively, there is tension on the value of $k$  from the radiation and matter simulations. We will discuss this further in the next section.

We can compare our results with the values of the parameters previously given in \cite{Martins:2008zz}, where they used two different sets of simulations to determine the value of their parameters. On the one hand they use the results  due to the work by Yamaguchi \cite{Yamaguchi:2001xn}, to obtain both the values of $\epsilon$ and the velocity:
\begin{eqnarray}
&&\epsilon_r \sim 1.3\pm 0.4\,,  \qquad c_r\sim1.3\pm 0.7\,, \nonumber\\
&&\epsilon_m \sim 1.6 \pm 0.1\,, \qquad c_m \sim 1.2\pm0.6\,.
\end{eqnarray}
On the other, they use the values of $\epsilon$ from the work by Bennett and Rhie \cite{Bennett:1990xy},  combined with the velocities obtained by Yamaguchi (Bennett and Rhie do not give estimates for the velocities):
\begin{eqnarray}
&&\epsilon_r \sim 1.3\pm0.2\,, \qquad c_r\sim1.3 \pm 0.5 \,, \nonumber\\
&&\epsilon_m \sim 1.9\pm 0.2\,,\qquad c_m \sim 1.4 \pm 0.7\,.
\end{eqnarray}

The uncertainties in $v_0$ were such that it was not possible to determine  whether monopoles move subluminally or luminally. Likewise, due to the uncertainties it was not possible to determine the $k$ parameter (from Eq.(\ref{eq-pa}))   in  \cite{Martins:2008zz}.

\section{Conclusions and discussion\label{co}}

In this article we have obtained the most accurate values of the average network velocity of a network of global monopoles to date. These values have then been used to complete the characterization of the phenomenological "velocity--dependent one-scale"  (VOS) model proposed by  \cite{Martins:2008zz}. This model admits two branches of scaling solutions, one with $v_0 =1 $ (luminal) and one with $v_0<1$ (subluminal).  The  velocities measured in our work make  it possible to determine, for the first time, that only the subluminal branch is physically realized.

In order to obtain the velocities, we have implemented a new method (the monopole--tracking method) to measure global monopole velocities in a network. This method has two main steps: first, we translate from field values to monopole positions, i.e., we calculate topological charge in field space, and translate it into a position in space. The second step identifies monopoles in each time slice, thus following the evolution of each monopole in time. This last step can be applied to any evolution of point like particles in the lattice, so our method is not specific for global monopoles and it can be used in many other situations.

The monopole--tracking  method has been used to calculate the average velocities of global monopoles in a network. Nevertheless, since we can obtain the behaviour of each one of the monopoles in the box throughout the whole evolution, this method can also be used to analyse and understand the complex mechanism governing  annihilation (and choice of annihilating partner) in global monopole models. We have observed how a monopole travels to meet with an antimonopole, but this antimonopole annihilates with a third monopole before the first one reaches it. Then, the monopole slows down, changes direction, and starts speeding up towards another monopole. These peculiar behaviours are interesting and are left for future work.

We have also implemented two other methods previously proposed in the literature.   One of them   was introduced to measure global monopoles using local variables in  \cite{Yamaguchi:2001xn}, and we call it Local Velocity Estimator (LVE). It was known that that method was prone to have high errors.  We have improved the approach, as well as perform simulations in bigger lattices, to obtain results with more reasonable errors. The other method was proposed in \cite{Hindmarsh:2008dw} (the Average Velocity Field Estimator (AVFE)) for the case of cosmic strings and uses weighted averages of field quantities to estimate directly the average network velocity. We have adapted it for the case of global monopoles and we used it in our simulations.

It is interesting to compare these two types of methods. The monopole--tracking  method follows the position of the monopole over their evolution; it is a method in spacetime. The other methods, the LVE and AVFE, extract the information from field-space at every time-step. These two approaches are in principle very different, but we have showed that 
the results coming from LVE and AVFE agree with the results obtained using our new monopole--tracking method.

Actually, the (improved) LVE method   agrees surprisingly well with the results obtained from our methods. The other method, the AVFE,   also agrees, but only within 1-$\sigma$ or 2-$\sigma$. The differences come from systematic errors that can be understood from physical considerations. For example, regions that contribute to the velocity estimation in AVFE may not be from monopole position, but can be from the remnants of a monopole-antimonopole annihilation.
In any case, this is a good test to show that the approaches made by considering field-theory information, and which are much easier to implement,   work well.

We were also able to compare the results obtained using the so-called   Press-Ryden-Spergel algorithm \cite{Press:1989yh} with the {\it true} simulations for the case of global monopoles. The former  allows for a rather larger dynamical range, but does not evolve the {\it true} equations of motion; instead it evolves some artificially modified equations. In the latter one has to pay the price of a short dynamical range for the benefit of simulating the true equations of motion. We show that in both cases the results obtained are compatible. This result could be extrapolated to cases where unfortunately the true equations of motion cannot be solved, and gives some reassurance that the Press-Ryden-Spergel algorithm is a  reasonable approximation.

Finally,  with the data obtained from the different velocity estimations, we obtain the average network velocity for global monopoles to be (\ref{vels})
$$v_r=0.76\pm0.07\,,\quad v_m=0.65\pm0.08\,, $$
for radiation and matter domination epochs, respectively.

We can compare the results obtained in this article with the velocities obtained   in the literature. For example, Yamaguchi \cite{Yamaguchi:2001xn} used the LVE to obtain the  velocity values for global monopoles as:
\be
v_r=1.0\pm0.3\,, \qquad v_m=0.8\pm0.3\,,
\ee
where $r$ stands for radiation era and $m$ for matter era. Our results are compatible with those of Yamaguchi's, but with much smaller errors. Note how Yamaguchi's numbers were not accurate enough to discard the ultrarelativistic branch.
 
The fact that we  get such an improvement on the velocity estimation has many reasons. Our monopole--tracking method is much more accurate than those previously in the literature. Also, our simulation is much bigger than those previously used. Besides, we also use combined the monopole--tracking methods with an improved version of the LVE and with the AVFE. Actually, had we used just the improved LVE method, we would have got much smaller errors than in the original work by Yamaguchi, since  the  improved version tries to minimize the known sources for errors, and   also, as mentioned above, because we used bigger simulations, increasing the statistical significance.

Our new velocity estimations allow us to determine that the physical branch of solutions of  the analytic model presented in \cite{Martins:2008zz} is the subluminal one. Actually, we have not found any monopole going close to  or above the speed of light in our simulations.
We can use our velocity  numbers, together with the network scaling numbers reported earlier,   to calibrate the analytic model  for global monopole networks in radiation and matter domination, respectively (see Table~\ref{tab-eps}):
\begin{eqnarray}
&&  c_r= 0.9 \pm 0.2\,, \qquad k_r= 0.9 \pm 0.1\,, \nonumber\\
&&  c_m= 0.7 \pm 0.2\,, \qquad  k_m= 1.6 \pm 0.2\,.
 \end{eqnarray}
The values of $c$   are compatible with those previously reported, but the errors have been reduced. The values of $k$ were never reported before, due to the uncertainties in previous simulations to determine whether the subluminal or the luminal branch was the one preferred  by monopoles.

Maybe more interestingly, the values of $c$ we obtain for radiation and matter domination are compatible with each other, even with the reduced error bars. However, there is tension on the values of $k$ from radiation and matter domination simulations. Already in the original paper for global monopole VOS model \cite{Martins:2008zz} the authors  comment on the different physics involved in the approximations, and how the term involving $c$ is more robust than the one involving $k$. Our simulations support their conclusions.

The reason why the values of $k$ depend on the simulation may be due to the approximations about the force between monopoles. The VOS model supposes the field configuration of a monopole to be spherically symmetric, and thus the mass of the monopole to grow linearly with distance. The model also asumes that the  force between monopoles is independent of distance. Moreover, the presence of other monopoles in the vicinity is factored in by adding and {\it  ad hoc} $1 / \sqrt {\cal N}$ factor. These assumptions should be revisited: the typical field configuration of a monopole in a network is unlikely spherical; the force between monopoles is independent of distance only if the monopoles are far from each other, it is not clear what the force is when the cores are involved; and the fact that there are other monopoles and antimonopoles distort the field configuration substantially. Our results point at a possible direction where the VOS model could be improved. In any case, one should also be cautious and bear in mind that there are different numerical errors in the simulations: there are errors due to the Press-Ryden-Spergel algorithm used, due to our algorithms to detect and estimate velocities, or even due to the inherent errors of the discretization.

The method described in this paper can be used also to characterize other types of defect networks. For example, a direct application will be to calibrate 
the  analytic models  for semilocal strings, where the knowledge obtained in this work about the treatment of the evolution of point like particles will be invaluable to track the velocities of the string ends.

\acknowledgments

We would like to thank Mark Hindmarsh, Carlos Martins  and Charlotte van Hulse for useful discussions.  We also would like to thank L. Sousa and P.P. Avelino for pointing out the error that was corrected in the Erratum.This work has been possible thanks to the computing infrastructure
of the i2Basque academic network, the COSMOS Consortium supercomputer (within the
DiRAC Facility jointly funded by STFC and the Large Facilities Capital Fund of BIS), and
the Andromeda/Baobab cluster of the University of Geneva. The authors acknowledge support from the Basque Government (IT-979-16), the Spanish Ministry MINECO (FPA2015-64041-C2-1P) and the Consolider Ingenio Programme EPI (CSD2010-00064). A.L-E. is also supported by the Basque Government grant BFI-2012-228.  AA's research was supported by a grant from the Simons Foundation, by the Foundation for Fundamental Research
on Matter (FOM) and the Netherlands Organization for Scientific Research (NWO/OCW).

\bibliography{monopoles}

\section*{Erratum}

In this paper  we obtain several velocity estimations for global monopoles. However, as is pointed out in \cite{Sousa:2017wvx}, there is an error in equation (5.4), which should read as follows:
\be
\epsilon=\frac{c v_0}{3(1-\lambda)-\lambda v_0^2}, \;\;\; k= \frac{\lambda v_0}{(1-\lambda)^{3/2}\epsilon^{1/2}}.
\ee

In Table~\ref{tab-an} and Table~\ref{tab-eps} we show the updated values of the corresponding Table 5 and Table 6 of the main paper.

\begin{table*}[!htb]
\begin{center}
\begin{tabular}{ |l| c  | c|c|c|c|c|}
\cline{2-7}
 \multicolumn{1}{c|}{}  & \multicolumn{3}{|c|}{s=0}  & \multicolumn{3}{|c|}{s=1}\\  \cline{2-7}
\multicolumn{1}{c|}{} & $\epsilon$ & c &k & $\epsilon$ & c & k \\ \hline
 Radiation  & 1.42$\pm$ 0.09& 2.5 $\pm$ 0.2 & 0.76 $\pm$ 0.02 & 1.53 $\pm$ 0.04& 2.6 $\pm$ 0.2 & 0.92 $\pm$ 0.02 \\ \hline
Matter & 1.97 $\pm$0.09 & 2.2 $\pm$ 0.2 & 1.55 $\pm$ 0.04 & 2.00 $\pm$0.06 & 2.7 $\pm$ 0.2 & 1.42$\pm$ 0.02 \\ \hline
\end{tabular}
\caption{\label{tab-an} Values of the analytic parameters for  radiation ($\lambda = 1/2$) and matter ($\lambda = 2/3$), and for $s=0$ and $s=1$.
}
\end{center}
\end{table*}

\begin{table*}[h]
\begin{center}
\begin{tabular}{ |l| c  | c|c|c|}
\cline{2-4}
\multicolumn{1}{c|}{} & $\epsilon$ & c &k  \\ \hline
 Radiation  & 1.47 $\pm$ 0.09& 2.6 $\pm$ 0.3 & 0.9 $\pm$ 0.1  \\ \hline
Matter & 1.98 $\pm$ 0.07 & 2.5 $\pm$ 0.3 & 1.6 $\pm$ 0.2 \\ \hline
\end{tabular}
\caption{\label{tab-eps} Values of the analytic parameters for radiation and  matter  averaging over all simulations with $s=0$ and $s=1$.  We first average over all velocities, and then use that average (with errors)  to obtain the value of $c$ and $k$.}
\end{center}
\end{table*} 

Finally the updated version of equation (6.2) reads as follows:
\be\nonumber
c_r = 2.6 \pm 0.3, \;\;\; k_r=0.9 \pm 0.1,
\ee
\be
c_m = 2.5 \pm 0.3, \;\;\; k_m=1.6 \pm 0.1.
\ee

These corrections do  not affect any of the results or conclusions that were obtained in our paper.

\end{document}